\documentclass{article}

    \PassOptionsToPackage{numbers, compress}{natbib}

    \usepackage[final]{tackling_climate_workshop_style}

\usepackage[utf8]{inputenc} 
\usepackage[T1]{fontenc}    
\usepackage{hyperref}       
\usepackage{url}            
\usepackage{booktabs}       
\usepackage{amsfonts}       
\usepackage{nicefrac}       
\usepackage{microtype}      

\usepackage{wrapfig}
\usepackage{graphicx}
\usepackage{esvect}
\usepackage{amsmath}
\usepackage{algorithm}
\usepackage{algpseudocode}
\newcommand{\cref}[1]{Chapter~\ref{#1}}  

\newcommand{\supplementarysection}{%
  \setcounter{figure}{0}
  \let\oldthefigure\thefigure
  \renewcommand{\thefigure}{S\oldthefigure}
  \let\oldchapter\chapter
  \renewcommand{\chapter}{
    \let\thefigure\oldthefigure
    \let\chapter\oldchapter
    \oldchapter
  }
}

\title{An Interpretable Model of Climate Change\\Using Correlative Learning}

\author{Charles Anderson \& Jason Stock\\
Computer Science\\
Colorado State University\\
\texttt{\{anderson,stock\}@colostate.edu}
}

\begin{document}

\maketitle

\begin{abstract}
Determining changes in global temperature and precipitation that may indicate climate change is complicated by annual variations.  One approach for finding potential climate change indicators is to train a model that predicts the year from annual means of global temperatures and precipitations. Such data is available from the CMIP6 ensemble of simulations. Here a two-hidden-layer neural network trained on this data successfully predicts the year. Differences among temperature and precipitation patterns for which the model predicts specific years reveal changes through time. To find these optimal patterns, a new way of interpreting what the neural network has learned is explored.  Alopex, a stochastic correlative learning algorithm, is used to find optimal temperature and precipitation maps that best predict a given year.  These maps are compared over multiple years to show how temperature and precipitations patterns indicative of each year change over time.
\end{abstract}

\section{Introduction}

Deep networks have been used successfully to model many complex
relationships in data from a wide variety of domains.  Both practice
and theory suggest that large, deep networks may generalize better to
untrained data samples. This leads to important questions regarding the
interpretability of such large networks to understand the limitations
and trustworthiness of these models.
In many studies involving measurements of the natural world, domain
experts are most familiar with simple statistical analysis, such as
linear regression.  Therefore, to simplify the interpretability and trust of
models, models that are simple extensions of linear models should be the first step.

Here we follow this suggestion in an attempt to find indicators of climate change using a simple two-hidden layer neural network trained on global temperatures and precipitations from the sixth phase of the Coupled Model Intercomparison
Project~\cite{Eyring16} (CMIP6).  The potential of discovering indicators of climate change from the previous CMIP5 data was demonstrated by Barnes, et al., \cite{Barnes19,Barnes20} who found that linear models map global temperature data to year quite well and a simplified interpretation of the models was performed by analyzing the resulting linear model weights. In follow-on work, they recast the regression
problem as a classification task in order to use Layerwise
Relevance Propagation (LRP) to identify spatial patterns significant
to classifying specific decades~\cite{Toms20}.
Here we return to the regression approach and  extend it by including temperature and
precipitation CMIP6 data and modeling it with a neural network having two hidden layers.  A correlative learning algorithm is then applied to interpret what our neural network has learned about how temperatures and precipitations change over the years.

In Section~\ref{sec:interpret}, a brief summary of local and global methods for interpreting what a neural network has learned are reviewed. Alopex is described as a global method. This is followed by a demonstration of the Alopex method by using it to interpret what a neural network has learned when trained on the MNIST data.  This same approach is then applied in Section~\ref{sec:results} to a neural network trained to predict the year from global temperature and precipitation data, revealing intriguing changes in this data over time.


\section{Interpretation of Neural Network Models} \label{sec:interpret}

\begin{figure}[t!]
    \centering
    \includegraphics[width=\linewidth]{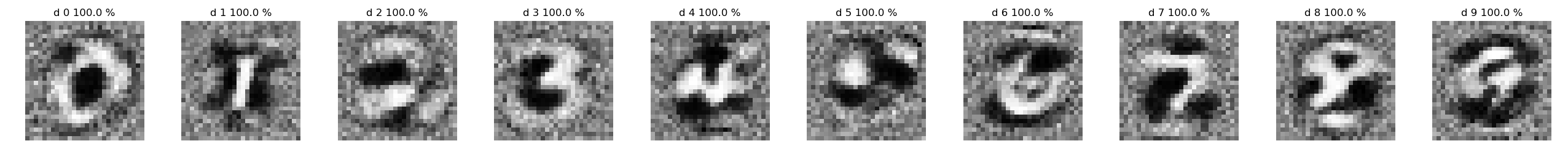}
    \caption{Input images generated by Alopex that maximize the likelihood of each digit. The images are averages of the final images from 20 repetitions of the Alopex algorithm.}
    \label{fig:mnist_images}
\end{figure}

Interpretability methods aim to convey information about a network's learning process and decision making. These methods largely fall in two categories, namely local and global. Local explanations provide insight on individual predictions. For example, saliency maps show certain pixels for a given sample with a positive saliency that increase accuracy in the prediction \cite{simonyan2013deep}. Further, methods such as LRP \cite{bach2015pixel}, DeepLift \cite{shrikumar2017learning}, and Gradient*Input \cite{shrikumar2016not} aim to provide pixel-wise relevance. In many cases these methods are mathematically equivalent (i.e., network's with ReLU activations or zero valued baselines) \cite{ancona2017towards} and are cumbersome to reason about over a large set of data.

Global methods describe the general behavior of a network where explanations are made on a class of predictions or network components. Most common for neural networks are methods to understand high-level concepts (e.g., color or texture sensitivity) \cite{kim2018interpretability}, visualizing maximal neuron or layer activations \cite{zeiler2014visualizing}, or optimizing an input to maximize the probability of an output \cite{simonyan2013deep}. In this work, we focus on finding an optimal input that generalizes to individual outputs. 

Both local and global methods that are gradient-free do exist (e.g., computing surrogate models as done with LIME \cite{ribeiro2016should} or Shapley value-based feature importance \cite{lundberg2017unified}). However, the majority of these methods rely on backpropagating gradients through the network (including \cite{zeiler2014visualizing,simonyan2013deep}). Here we introduce Alopex as a global interpretaton method that does not rely on gradients. Specifically, we leverage local correlations between changes in input features and changes in the global error function to produce a global view on a class of predictions.

\begin{wrapfigure}{r}{0.5\textwidth}
  \begin{center}
   \begin{minipage}{\linewidth}
   \vspace{-2\baselineskip}
    \begin{algorithm}[H]
        \caption{Alopex}\label{alg:alopex}
        \begin{algorithmic}[1]
                \State $f(x)\;\text{is neural network model}$
                \State $T \gets 1.0$
                \State $K \gets 100,000$
                \State $\delta \gets 0.0002$
                \State $m_0 \gets 0$
                \State $t \gets \text{digit}\text{ or year}$
                \State $x_0 \gets \text{vector of constant midrange values}$
                \For {$k \in \{1,...,K\}$}
                    \State $y_k \gets f(x_k)$
                    \State $e_k \gets loss(y_k, t)$
                    \State $c_k \gets (e_k - e_{k-1}) (x_k - x_{k-1})$
                    \State $p_k \gets  1 / (1 + e^{-c_k/T})$
                    \State $d_k \gets \begin{cases}
                                        \;\;\delta, & \text{w.p. } p_k\\
                                        -\delta, & \text{w.p. } 1-p_k
                                        \end{cases}$
                    \State $x_{k+1} \gets x_k + d_k + \lambda m_k$
                    \State $ m_{k+1} \gets d_k + \lambda m_k$
                    \State $T_{k+1} \gets 0.9998 T_k$
                \EndFor
                \State \Return $x_{k+1}$

        \end{algorithmic}
    \end{algorithm}
       \vspace{-2\baselineskip}
 \end{minipage} 
  \end{center}
\end{wrapfigure}

The Alopex~\cite{Harth74} algorithm was developed by E. Harth to investigate the receptive field of neurons in the frog optic tectum.  The intensities of individual pixels in an image presented to the frog were randomly initialized and updated during a number of steps.  In each step the correlation between changes in a pixel's intensity and changes in the neuron's firing rate determined which way to adjust each intensity, which was increased if the correlation was positive and decreased if it was negative. This simple correlative learning process was adapted by K.P. Unnikrishnan, et al.,~\cite{unni94} to train feedforward and recurrent neural networks.  In a very similar manner to Harth's original work, we use Alopex to incrementally adjust the input values to our neural network using the correlations between input changes and changes in the loss of our model.  This process converges on images that maximally predict a certain class in a classification problem, or a certain output value in a regression problem.  The Alopex algorithm is summarized in Algorithm~\ref{alg:alopex}, where $x$ is the input to our neural network model, $f(x)$ is the output of the model and $loss(y_k, t)$ is the loss for predicted value $y_k$ and correct target value $t$.

Alopex is demonstrated by applying it to a simple network trained on the MNIST digits. The network contains two layers of 20 convolutional units each with kernel size 5 x 5 and stride of 1 x 1.  The network is trained using M{\o}ller's Scaled Conjugate Gradient Algorithm~\cite{Moller93} with cross-entropy loss. For each digit, Alopex is applied 20 times starting with different initial pixel values. The mean of the 20 resulting images are shown in Figure~\ref{fig:mnist_images}. All 20 images for each digit were classified correctly.  Since Alopex starts with randomly-initialized pixel values, it is not dependent on particular image samples. Thus, this global method converges on images that are most confidently predicted as coming from the correct class. White, or high intensity, pixels strongly correlate with the digit class, while black pixels negatively correlate with the digit class.  Pixels near the edges are medium intensity showing their values have little correlation with the digit class. Many questions arise when studying these images and they are being investigated in on-going work.  Here we use MNIST just as a demonstration.

\section{Results on Climate Change Modeling} \label{sec:results}

\begin{figure}
    \centering
    \begin{tabular}{cc}
    \includegraphics[width=0.6\linewidth]{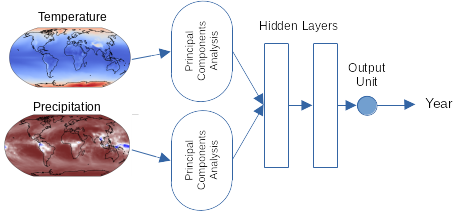}
    &
    \includegraphics[width=0.4\linewidth]{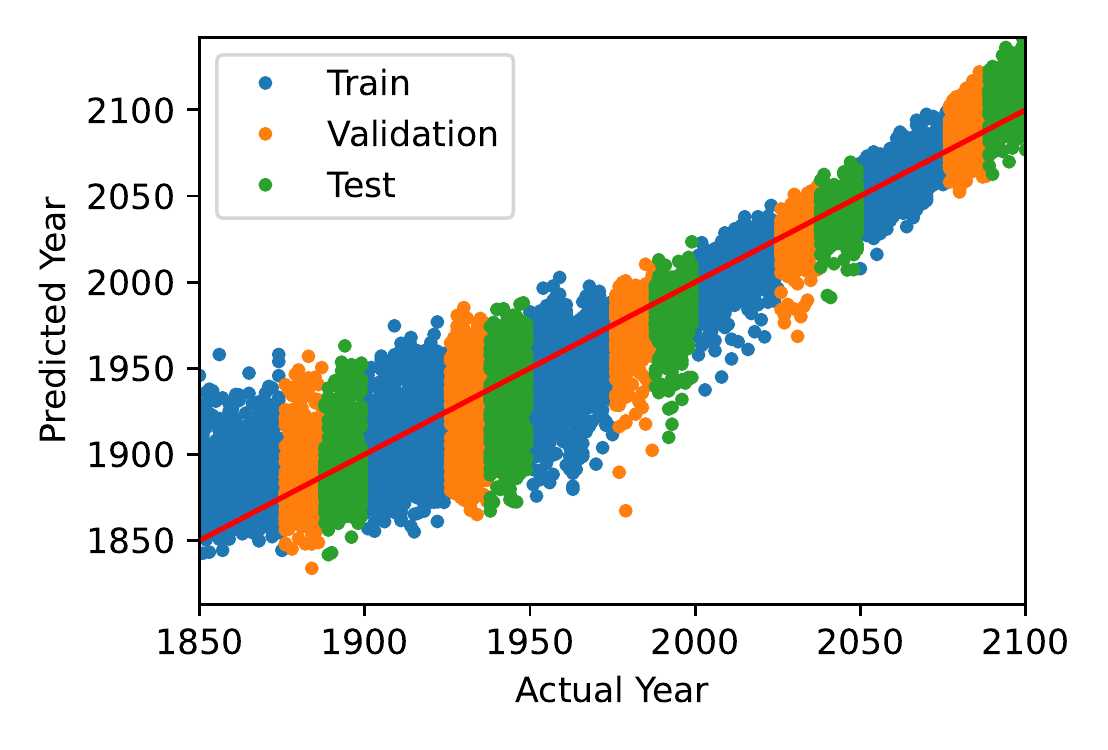} \\ 
    a. & b.
    \end{tabular}
    \caption{a. Model structure with 40 tanh units in each of two hidden layers and a linear output unit. b. Predicted versus actual year for all samples, which were partitioned into train, validation, and test subsets by years.}
    \label{fig:cmip-net}
\end{figure}

The CMIP6~\cite{Eyring16} data was produced by 35 models of earth's
atmosphere from which simulated global temperature and precipitation
maps can be obtained for years 1850 to 2100. This data has a spatial resolution of 120 latitude and 240
longitude values. Not surprisingly, there are numerous correlations
among temperatures and precipitations at multiple spatial locations, which was dealt with in prior work by
performing ridge regression to limit the magnitude of weights in the
first layer of the neural network models~\cite{Barnes19}.  Here we use
an alternative approach, Principal Components Analysis (PCA), to represent the data with independent factors and to decrease the data dimensionality by projecting the $2\times 120\times 240$ or $57,600$ dimensional annual samples to the first 250 singular vectors, i.e., the ones capturing most of the variance in the data. The choice of 250 singular vectors was made by comparing the RMSE in predicted year for the validation data set when using the different quantities of top singular vectors from 1 to 1000. 

A two-layer fully-connected neural network was trained on the CMIP6 data as follows.  Each hidden layer contained 20 tanh units. PCA was applied separately to the temperature and precipitation data and the resulting 250-dimensional vector was input to the first hidden layer. The output layer of the neural network was a single linear unit trained to predict the year.  Figure~\ref{fig:cmip-net}a shows the model structure and Figure~\ref{fig:cmip-net}b shows the results of training the network using the Scaled Conjugate Gradient algorithm~\cite{Moller93} to minimize the mean square error in the predicted year. This plot suggests that years after about 2000 are easier to predict, possibly due to human-caused forcing functions included in the CMIP6 models.

\begin{figure}
    \centering
     \includegraphics[width=1.0\linewidth]{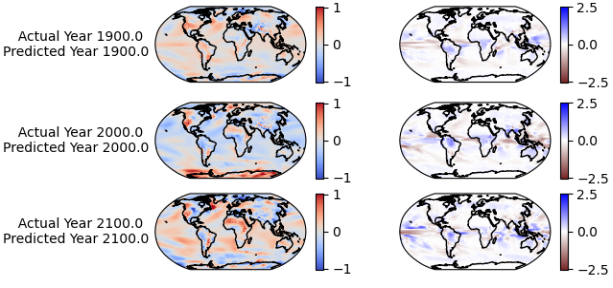}
    \caption{Patterns in global temperature and precipitation that optimally relate to particular years. Colors correspond to differences from the temperature and precipitation maps found for year 1850. The left column of maps show areas of higher temperatures than those for 1850 in red and lower temperatures in blue. The right column of maps show areas of higher precipitation levels than those in year 1850 in blue and lower values in brown.}
    \label{fig:cmip-three-maps}
\end{figure}

Similar to the above demonstration of Alopex on the MNIST data, we applied Alopex to the CMIP6 model by searching for input temperature and precipitation maps that best predicted particular years.  Alopex was initialized with PCA projected values near their median values with uniformly-distributed noise added.  This was repeated 20 times. For each result the PCA projection was inverted to produce temperature and precipitation maps and the means of the resulting maps were calculated. The resulting maps for year 1850 were subtracted from the maps for following years to highlight changes from 1850. Figure~\ref{fig:cmip-three-maps} illustrates this result for target years 1900, 2000, and 2100. Results for other years are shown in Figure~\ref{figsupp:cmip-maps}.

Several interesting patterns can be discerned in these images.  Year 2000 shows considerable warming in the Antarctic region and western United states.  Year 2100 shows warmer temperatures in many areas. The precipitation maps on the right side show drier regions along the Pacific equator in Years 1900 and 2000. In Year 2100 wetter areas north and south of the equator in the Pacific Ocean are apparent. Similar questions arise when considering maps for other years as shown in Supplementary Figure~\ref{figsupp:cmip-maps}.

\section{Conclusion}

Results shown here suggest that the correlative learning algorithm, Alopex, may be helpful in interpreting neural network models by finding global, approximately optimal, input patterns corresponding to particular network output values.  It can be similarly used to find global input patterns that cause any unit in the neural network to respond maximally. The fact that it is a correlative learning algorithm means that gradients are not required.  However, it would be interesting to compare the results obtained here with other interpretation methods, especially gradient-based global approaches.

The temperature and precipitation maps resulting from the Alopex method reveal interesting changes over the years. Currently, specialists in climate and atmosphere modeling are being consulted to assist in determining relationships between these patterns and changes that are expected from current knowledge of effects on climate due to human activities.

\acksection
This work is supported by NSF Grant No. 2019758, \textit{AI Institute for Research on Trustworthy AI in Weather, Climate, and Coastal Oceanography (AI2ES)}.

\bibliography{main}


\newpage
\section*{Appendix}

\begin{figure}[!h]
    \centering
    \includegraphics[width=\linewidth]{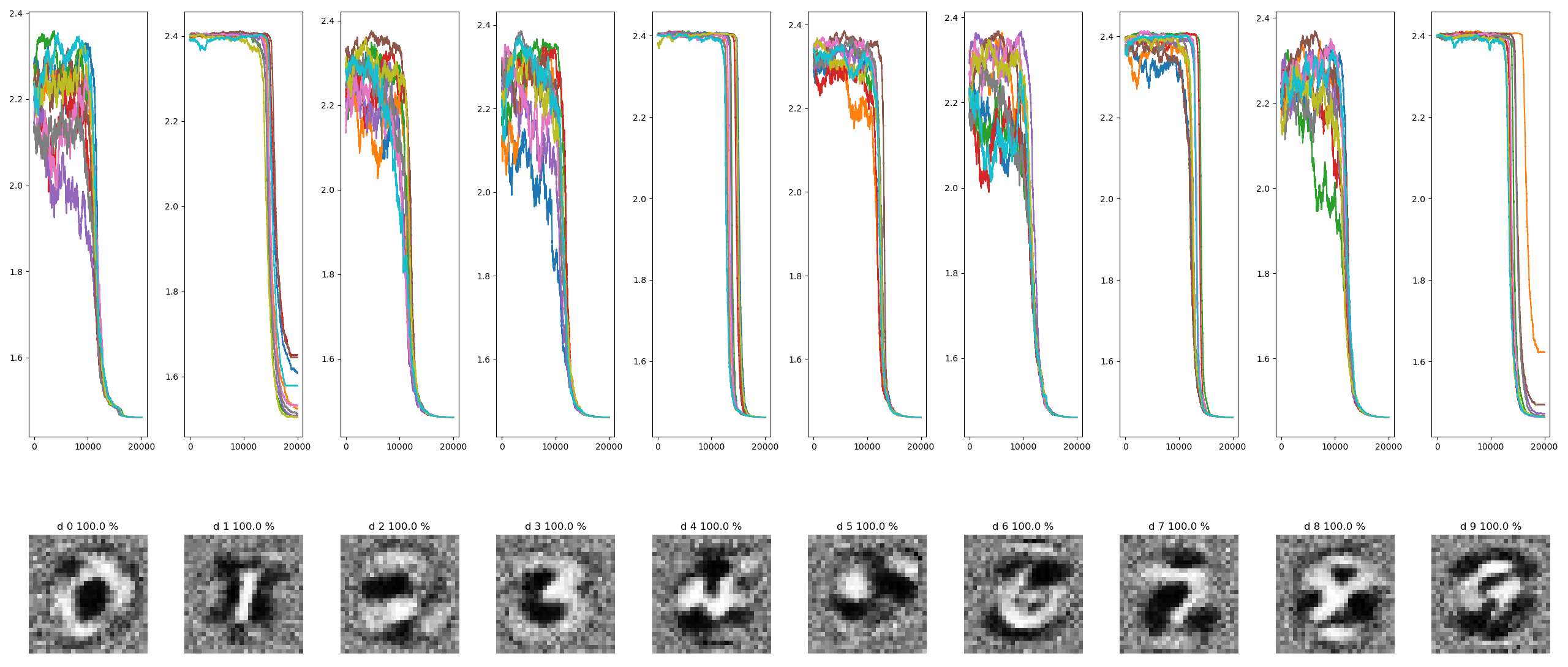}
    \caption{Input images generated by Alopex that maximize the likelihood of each digit. Above each image are graphs of the loss versus iterations of the Alopex algorithm. The images are averages from 20 repetitions of the Alopex algorithm. The loss of each iteration is plotted above each image, with different colors for different repetitions.}
    \label{figsupp:mnist_images_and_loss}
\end{figure}

\begin{figure}[!h]
    \centering
    \includegraphics[width=\linewidth]{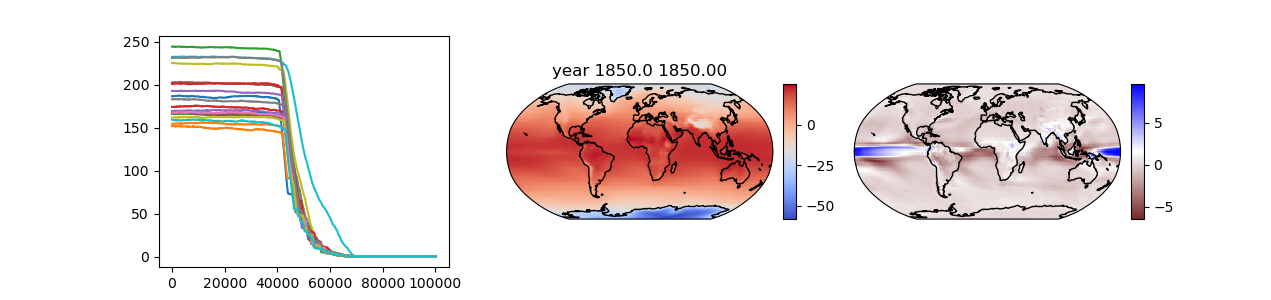}\\
     \includegraphics[width=\linewidth]{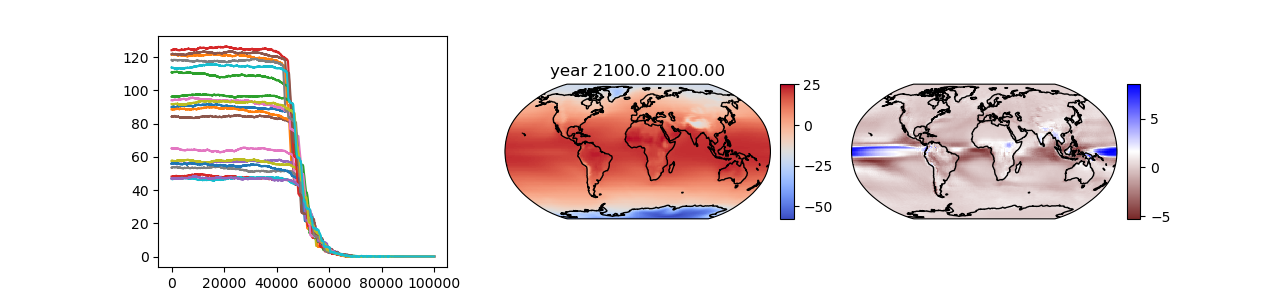}\\
    \caption{Examples of Alopex applied to the CMIP6 neural network to generate temperature and precipitation maps that most confidently predict Year 1850 (top row) and 2100 (bottom row). On the left are plots of the mean square error in predicted year verus steps of the Alopex algorithm. Different colors are for 20 different initializations of Alopex.}
    \label{figsupp:cmip_maps_and_loss}
\end{figure}

\begin{figure}
    \centering
     \includegraphics[height=0.9\textheight]{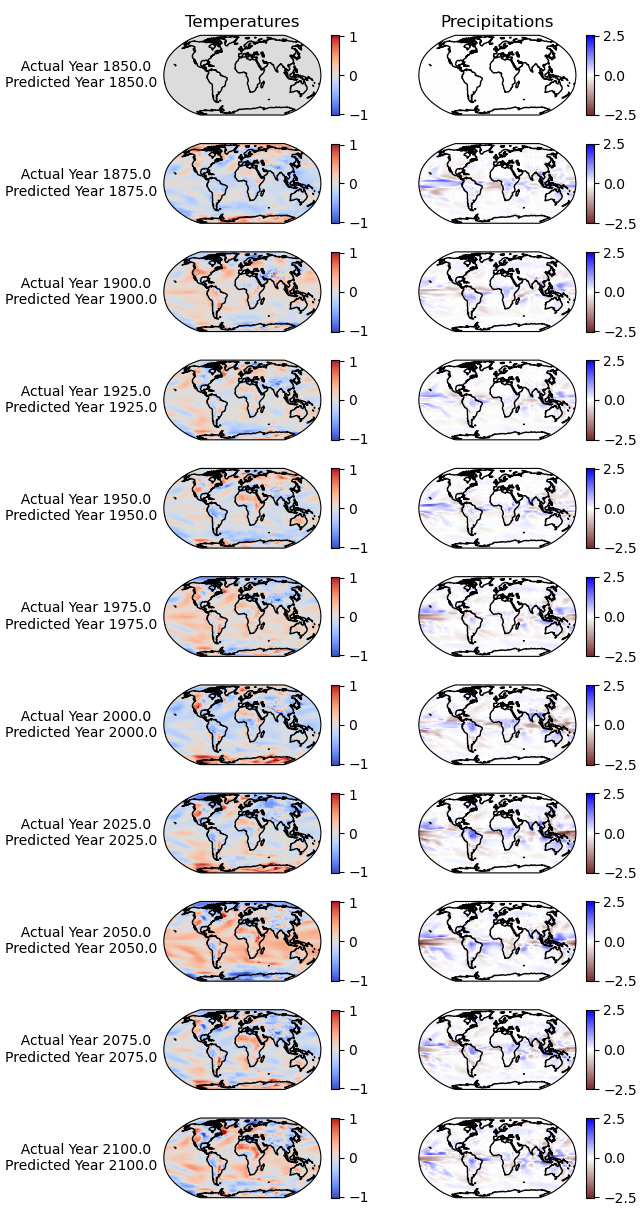}
    \caption{Patterns in global temperature and precipitation that optimally relate to a sequence of years. Colors correspond to differences from the mean temperatures and precipitations. The left column of maps show areas of higher temperatures than the mean in red and lower temperatures in blue. The right column of maps show areas of higher precipitation than the mean in blue and lower values in brown.}
    \label{figsupp:cmip-maps}
\end{figure}

This appendix includes additional figures illustrating the convergence of the Alopex algorithm for the MNIST digits (Figure~\ref{figsupp:mnist_images_and_loss}) and for two years of the CMIP6 data (Figure~\ref{figsupp:cmip_maps_and_loss}).  It also includes temperature and precipitation maps that most confidently predict years 1875, 1900, 1925, $\ldots$, 2100 as differences from the maps of 1850 (Figure~\ref{figsupp:cmip-maps}).

\end{document}